\documentclass[journal,final,twocolumn,10pt,twoside]{IEEEtranTCOM}
\normalsize
\usepackage{amssymb}
\usepackage{amsmath}

\usepackage{graphicx}

\ifCLASSINFOpdf
\else
\fi

\usepackage{mdwmath}

\usepackage{gensymb}

\usepackage{eqparbox}

\usepackage{fixltx2e}

\usepackage{mdwmath}
\DeclareMathOperator{\E}{\mathbb{E}}

\usepackage{color}
\usepackage{xcolor}
\date{}
\begin{document}
\title{Cellular Underwater Wireless Optical CDMA Network: Potentials and Challenges
 } %
\author{\normalsize Farhad~Akhoundi \IEEEauthorrefmark{2},
Mohammad~Vahid~Jamali \IEEEauthorrefmark{2},
Navid~Bani Hassan \IEEEauthorrefmark{2},
Hamzeh~Beyranvand \IEEEauthorrefmark{3},
Amir~Minoofar \IEEEauthorrefmark{2},
        and~Jawad~A.~Salehi \IEEEauthorrefmark{2},~\IEEEmembership{\normalsize Fellow,~IEEE}
\thanks{\IEEEauthorrefmark{2} The authors are with the Optical Networks Research Laboratory (ONRL), Department of Electrical Engineering, Sharif University of Technology, Tehran, Iran (e-mail:akhoundi@alum.sharif.edu; mohammad.v.jamali@gmail.com; n.banihasan@modares.ac.ir; amir\_minoofar@ee.sharif.edu; jasalehi@sharif.edu).} 
%
\thanks{\IEEEauthorrefmark{3} The author is with Electrical Engineering Department, Amirkabir University of Technology, Tehran, Iran (email: beyranvand@aut.ac.ir).} }
\maketitle
\begin{abstract}
 \textbf{Underwater wireless optical communications is an emerging solution to the expanding demand for broadband links in oceans and seas. In this paper, a cellular underwater wireless optical code division multiple-access (UW-OCDMA) network is proposed to provide broadband links for commercial and military applications. The optical orthogonal codes (OOC) are employed as signature codes of underwater mobile users. Fundamental key aspects of the network such as its backhaul architecture, its potential applications and its design challenges are presented. In particular, the proposed network is used as infrastructure of centralized, decentralized and relay-assisted underwater sensor networks for high-speed real-time monitoring. Furthermore, a promising underwater localization and positioning scheme based on this cellular network is presented. Finally, probable design challenges such as cell edge coverage, blockage avoidance, power control and increasing the network capacity are addressed.}
\end{abstract}
\begin{IEEEkeywords}
Underwater wireless optical communications, optical CDMA networks, underwater sensor networks, relay-assisted transmission, MIMO, localization and positioning, power control.
\end{IEEEkeywords}
\section{Introduction}
Rapidly growing commercial and military applications for underwater communication demands for a reliable, flexible and practical multi-access network. Recent studies have shown that there are two major solutions to this demand: acoustic and optical transmission. Due to the limited bandwidth of acoustic systems, the maximum achievable rates are restricted to 10$\sim$100 kbps. Furthermore, the low speed of acoustic waves in undersea media ($\approx$1500 m/s) which results in a high latency in long range communications causes problems for synchronization and multiple access techniques. Alternatively, in comparison with the traditional acoustic approach, wireless optical communication has three main advantages: higher bandwidth, higher security and lower time latency.

Despite their promising advantages, limited attainable communication range of underwater wireless optical communication (UWOC) systems, \textit{i.e.}, less than 100 meters with realistic average transmit powers hampers their widespread usage. In general, optical beam propagation through water suffers from three main disturbing effects: absorption, scattering and turbulence. 
In \cite{mobley1994light}, Mobley has accomplished an in-depth study of light interaction in water to characterize absorption and scattering effects of different water types based on theoretical analysis and experimental evidence. Consequently, Tang \textit{et al.} succeed to present a closed-form expression of double Gamma function to model the channel impulse response in the presence of absorption and scattering effects \cite{tang2014impulse}. Optical channel turbulence is also studied, though in less extent, by Korotkova \textit{et al.} in \cite{korotkova2012light}. Although many research activities have been carried out to design and analyze point-to-point underwater optical communication link, lack of a promising multi-access optical underwater network is obvious in the literature.

This research is inspired by the need to design an underwater wireless optical network with multiple-access capability to extend the boundaries and make possible communication among various fixed and mobile users in a relatively large underwater area. This network can be exploited in a variety of applications such as imaging, real-time video transmission, high throughput sensor networks, and also can potentially provide reliable communication links for unmanned underwater vehicles (UUVs), submarines, ships, buoys, and in particular divers.

Among many multiple-access schemes, optical code division multiple access (OCDMA) is receiving much attention as a promising access technique to share common resources among asynchronous users without any central controller, which is highly desirable in underwater environment. Amongst the first generation OCDMA-based systems, using optical orthogonal codes (OOC) in fiber-optic communications was introduced by Salehi in 1989 \cite{salehi1989code}, while capability of this scheme to free space, infrared indoors and visible light communication have been recently studied \cite{salehi2007emerging}. Furthermore, performance analysis of an underwater wireless optical CDMA (UW-OCDMA) network is presented in \cite{akhoundi2014cellular}. In this article, we elaborate possible challenges and potential applications of cellular UW-OCDMA network based on OOCs. In a typical UW-OCDMA network, mobile and fixed users communicate to an optical base transceiver station (OBTS). Each active user transmits its data using a unique OOC code.

In particular, this article describes the proposed UW-OCDMA network architecture and discusses its potential application in local sensor networks and underwater localization.  Furthermore, possible challenges regarding blockage avoidance, cell edge coverage, power control algorithms, and increasing the number of active users are discussed.
\begin{figure*}
\centering
\includegraphics[width=7in]{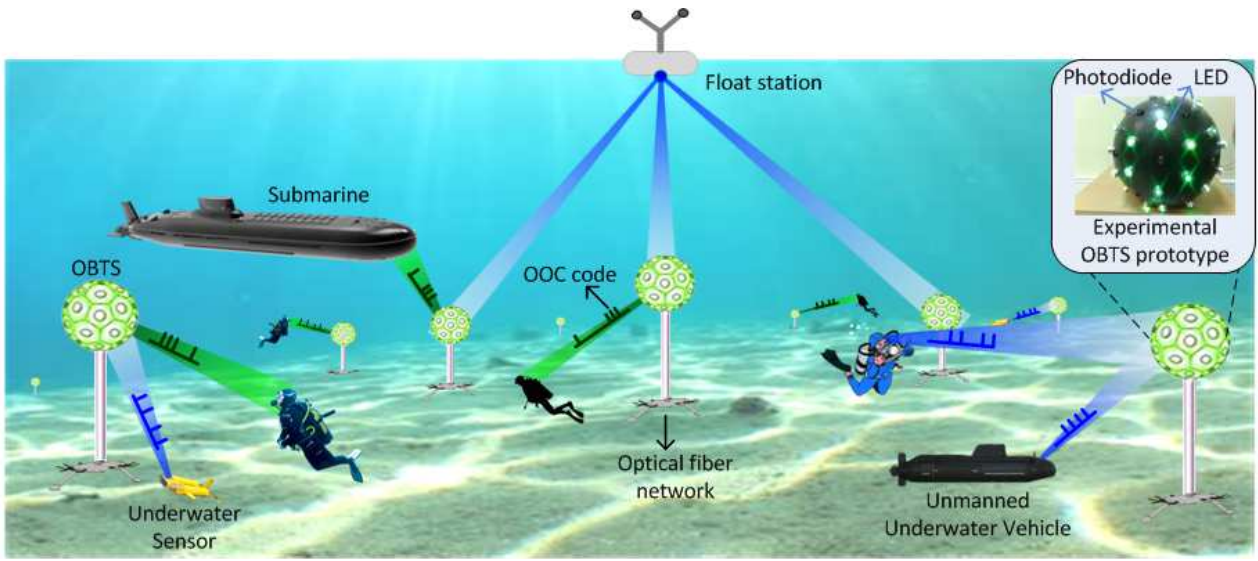}
\caption{The concept and geometry of the proposed cellular UW-OCDMA network with divers, submarines, and UUVs as mobile users.}	
\end{figure*}
\section{Cellular UW-OCDMA Network Architecture}
The general concept and geometry of the cellular UW-OCDMA network is illustrated in Fig. 1, in which a set of omnidirectional OBTSs are placed in the center of a hexagonal cell to cover a larger underwater area; and there are a number of user transceivers, each having a pair of optical unidirectional transmitter and omnidirectional receiver. All OBTSs are interconnected together via fiber optic network which can be linked to an optical network controller (ONC). As such, we can expand the limited underwater optical communication range to a relatively much larger area.
\subsection{System Model Description}
 
The shape of the designed OBTS resembles a soccer ball which is created from twelve regular pentagonal and twenty regular hexagonal panels positioned in a spherical geometry. We place $60$ LEDs on the junctions of this shape and $32$ photodetectors on the center of its pentagonal and hexagonal panels. Therefore, the OBTS acts as an omnidirectional transceiver. In regards to the density of active underwater users and their required data rate an appropriate set of OOC signatures is generated and each code is assigned to each user. Using its unique signature code, each user spreads or encodes its data at the transmitter. On-off keying (OOK) signaling is exploited to modulate users' data using intensity modulation (IM) technique.

Nearby OBTS receives and photodetects active underwater users' signals. Based on the network topology, the received signal will be either despread by OOC codes at the OBTS, or again converted back to an optical signal by applying an electro-optic Mach-Zehnder modulator; and then the optical signal will be transmitted to the ONC via fiber optic network. In other words, an OBTS can play the role of either a dummy received and forward or an intelligent decode and forward relay. A location database should be designed at the ONC to register the location of mobile users (MUs) and decide to which OBTS data must be forwarded. Backhaul architecture of the network will be discussed in more detail in subsection II-C.

An OOC is a sequence of $``0"$ and $``1"$, which is characterized by $(F,W,\rho)$ where $F$ is the code length, $W$ is the code weight which determines the total number of ones in each codeword, and $\rho$ is the maximum value of shifted autocorrelation and crosscorrelation \cite{salehi1989code}. The maximum number of OOC signatures ($N_c$) is limited by the well-known \textit{Johnson upper bound} \cite{salehi1989code} which restricts the total number of active MUs in our proposed cellular UW-OCDMA network. 

\subsection{Channel Model Description}
In regards to the impairing effects of underwater channel on the proposed UW-OCDMA network, three major phenomena namely absorption, scattering and turbulence need to be considered. Absorption is due to photon energy loss as a result of interaction with water molecules or other particulates via a thermal process while scattering is deviation of photons from their original path as a result of encountering photons with particulates. Loss in energy caused by absorption and scattering can be characterized by absorption coefficient $a(\lambda)$ and scattering coefficient $b(\lambda)$, respectively, with $\lambda$ denoting the optical wavelength. According to the Beer's law, non-scattered optical beam experiences an exponential extinction with cumulative coefficient  $c\left(\lambda \right)=a(\lambda )+\ b(\lambda )$. Optical turbulence, on the other hand, occurs due to the random variations of refractive index. These random variations in underwater medium mainly result from fluctuations in temperature and salinity and will cause fading on the received optical signal \cite{korotkova2012light}.

It is popular in the literature to model the absorption and scattering effects based on Monte Carlo (MC) simulation. This approach results in a fading-free impulse response $h_{0}(t)$. For instance in \cite{tang2014impulse}, Tang \textit{et al.} presented a closed-form expression of double Gamma functions to model the channel impulse response in the presence of absorption and scattering effects for coastal and harbor water environments where attenuation length $\tau=c(\lambda)L$ has relatively large values, in which $L$ denotes the link range. The closed-form expression of the double Gamma function is;
\begin{equation}
h_0(t)=C_1 \Delta te^{-C_2 \Delta t}+C_3  \Delta te^{-C_4  \Delta t},(t\geq t_0 )
\end{equation}
where $ \Delta t=t-t_0$, $t$ is the time scale and $t_0  = L/v$ is the propagation time which is the ratio of link range $L$ over light speed $v$ in water.
The parameter set $(C_1,C_2,C_3,C_4)$ in above equation can be computed from Monte Carlo simulation results using nonlinear least square criterion as;
\begin{equation}
\label{argmin}
〖(C〗_1,C_2,C_3,C_4)={\rm argmin}\left( \int〖〖[h_0(t)-h_{mc} (t)]〗^2 dt〗\right)
\end{equation}
where $h_0(t)$ is the double Gamma functions model in equation (\ref{argmin}),  $h_{mc}(t)$ is the Monte Carlo simulation results of impulse response, and ${\rm argmin(.)}$ is the operator to return the argument of the minimum \cite{tang2014impulse}.

 To characterize turbulence effects, the authors in \cite{jamali2015performance} have considered a positive multiplicative fading coefficient. In weak oceanic turbulence scenario, the fading coefficient can be modeled with log-normal distribution as;
\begin{equation}
f_{\tilde{h}}\left(\tilde{h}\right)=\frac{1}{2\tilde{h} \sqrt{2\pi {\sigma }^2_x}}{\rm exp}\left(-\frac{{\left({\ln  \left(\tilde{h}\right)\ }-2{\mu }_x\right)}^2}{8{\sigma }^2_x}\right) 
\end{equation}
where $\mu_x$ and $\sigma_x^2$ are, respectively, the mean and variance of the Gaussian distributed log-amplitude factor $x= \frac{1}{2} \ln(\tilde{h})$. To ensure that the fading coefficient conserves the energy, we normalize fading amplitude such that $\E[\tilde{h}]=1$, which implies $\mu_x=-\sigma_x^2$ . It can be shown that variance of log-amplitude factor $\sigma_x^2$ is related to the scintillation index of propagating signal $\sigma_I^2$ as $\sigma_x^2=\frac{1}{4} \ln⁡(\sigma_I^2+1)$. Therefore, having scintillation index $\sigma_x^2$ can be obtained for weak oceanic turbulence \cite{jamali2014analytical}. As a result the overall channel impulse response can be modeled as $h(t)=\tilde{h}h_{0}(t)$.

It has been shown in \cite{mobley1994light} that absorption and scattering have the lowest effect at the wavelength interval $400\ {\rm nm} < \lambda < 530\ {\rm nm}$ which span on the blue/green region of the visible light spectrum. In our proposed architecture, in order to reduce backscattered light of the OBTS's optical transmitters on its receivers we have chosen green LEDs with central wavelength of $532$ {\rm nm} for OBTS, blue LEDs with central wavelength of $450$ {\rm nm} for users, and appropriate optical filters to pass only the desirable range of wavelengths.


\begin{figure*}
 \centering
  \includegraphics[width=6in]{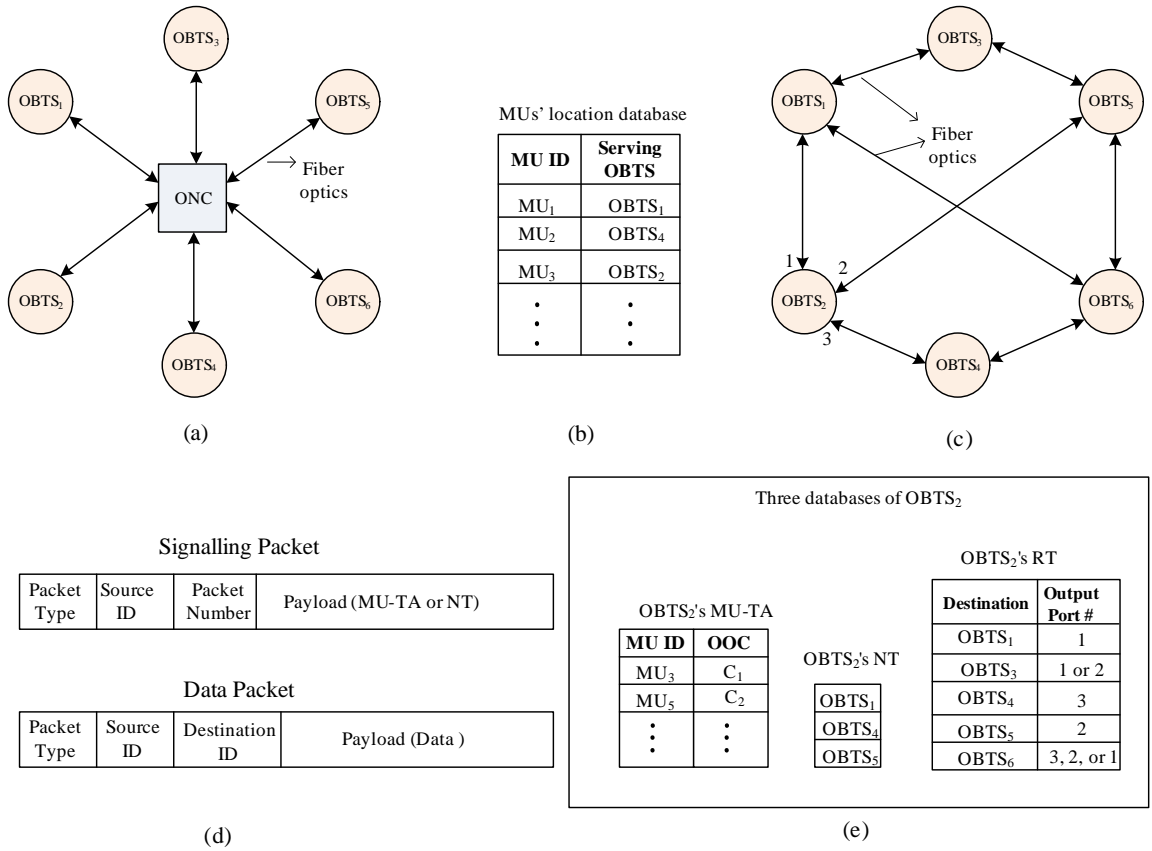}
  \caption{(a) An illustrative example of the centralized backhaul architecture; (b) MUs’ location database used in ONC; (c) an illustrative example of the decentralized backhaul architecture; (d) packet format in decentralized architecture; (e) databases corresponding to $\rm {OBTS}_2$ such as MU-TA, NT, and RT in the decentralized architecture.}
\end{figure*}
\subsection{Proposed Backhaul Architecture}
In the this subsection, the proposed backhaul architecture for underwater cellular network is discussed. Generally, two approaches are considered to interconnect OBTSs, namely centralized and decentralized. 
In centralized backhaul architecture, all OBTSs are connected to a central ONC, which is responsible to perform traffic forwarding among OBTSs. In Fig. 2(a), topology of the proposed architecture is illustrated.  In ONC database, the serving OBTS of each MU is recorded, thereby ONC interconnects MUs located in different OBTSs, as shown in Fig. 2(b). The ONC's database is updated based on the OBTS feedbacks after any changes in their MU association table (MU-AT). MU-AT contains the list of all MUs served by an OBTS and is updated after registration/elimination of MUs. 

In Fig. 2(c), the architecture of decentralized backhaul architecture is depicted. We note that in the decentralized backhaul architecture, each OBTS has an interface executing routing and traffic forwarding functionalities among OBTSs. Furthermore, a signaling protocol is used to distribute MU-ATs over the backhaul network. Each OBTS notifies its updated MU-AT to other OBTSs by broadcasting its MU-AT via a signaling packet. The header of the signaling packet contains packet type, source ID, and packet number, as depicted in Fig. 2(d). This packet is flooded in the network, \textit{i.e.}, each intermediate node broadcasts the received signaling packets to all its ports, thereby all OBTSs obtain the location information of MUs. It is worth noting that intermediate nodes record the packet number of signaling packets in order to prevent broadcasting a signaling packet two times and further to avoid unstable packet flooding.

\begin{figure*}
 \centering
  \includegraphics[width=6in]{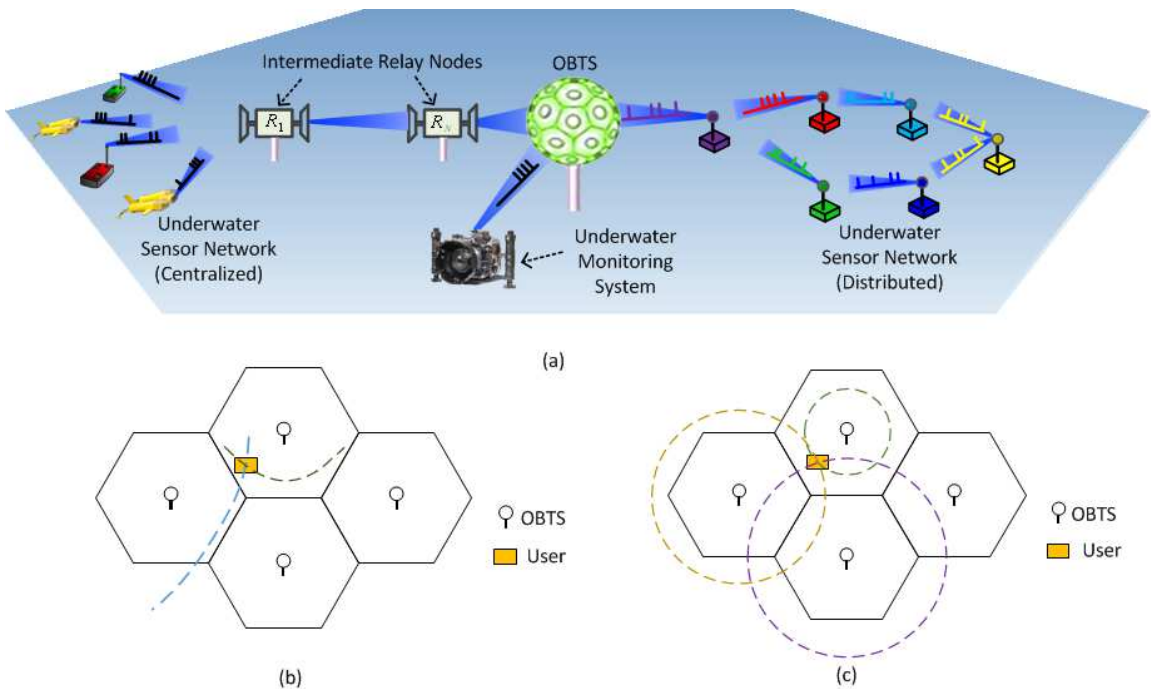}
  \caption{Potential applications of cellular UW-OCDMA network: (a) underwater optical sensor networks (centralized and distributed); (b) underwater localization based on TDOA scheme; (c) underwater localization based on RSS scheme.}
\end{figure*}
In decentralized architecture, in addition to database of MUs’ location, a routing table (RT) is necessary to send traffic between OBTSs. Each node discovers its neighbors by sending Hello packets, whereby neighbors respond the Hello packets by sending their OBTS ID. Then, information of neighbor OBTSs are recorded in neighboring table (NT). In order to discover the whole network topology each node broadcasts its NT periodically. By considering backhaul topology, each OBTS uses Dijkstra shortest path algorithm to compute its routing paths toward all other OBTSs, thereby its RT is completed by determining the output port number for each destination node.  Figure 2(e) illustrates different databases of $\rm OBTS_2$ in the decentralized architecture shown in Fig. 2(c).

\section{Potential Applications}
Besides the main purpose of the cellular UW-OCDMA network, which is providing a reliable and flexible  communication link for underwater mobile users in a relatively large area, there are yet many other potential applications which can exploit this infrastructure by adding minor complexity to the network. In the following subsections, two major secondary applications of the network are discussed.
\subsection{Underwater Optical Sensor Network}
Underwater sensor networks will find critical roles in investigating climate change, disaster prevention (e.g., tsunami), in monitoring biological, biogeochemical, evolutionary and ecological changes in the sea, ocean and lake environments, in pollution monitoring, and in helping to control and maintain oil production facilities. Underwater observation can be carried out with either sensors mounted on the see floor or UUVs equipped with sensors as shown in Fig. 3(a).

\subsubsection{Centralized underwater optical sensor networks}
In the centralized sensor network based on UW-OCDMA, each sensor employs its assigned OOC to encode and transmit its measured data to the nearby OBTS which may be located relatively far away from them. Communicating directly with the OBTS is practically impossible for these little battery-powered sensors. To overcome this issue, we can insert one or more relays in a serial configuration to collect sensors' data from a shorter distance. In other words, serial relaying is an attractive candidate for broadening communication coverage for limited-power transmitters. Furthermore, since degrading effects of absorption, scattering and turbulence rapidly increase with the communication distance, serial relaying or multi-hop transmission can be employed to considerably mitigate the channel impairments. Although different schemes such as decode-and-forward (DF) can be applied, to simplify relays structure, chip detect-and-forward (CDF) algorithm, which is a promising technique in OOC-based OCDMA systems, can be adopted in relay nodes. In this scheme, the relay node first decides on the presence or absence of each chip; and then forwards the detected chip to the OBTS for further analysis \cite{chu2014performance}. Note that the first relay $R_1$, has relatively wide field of view (FOV) to receive optical signals from all sensors; therefore, CDF process in this node is affected by multiple access interference (MAI). However, exploiting chip level detector with hard limiter can substantially reduce this interference \cite{zahedi2000analytical}. 

In order to analyze the end-to-end BER of relay-assisted UW-OCDMA network, we assume that  the transmitted bit is ``$1$" if all detected chips are ``ON" and otherwise we recognize ``$0$" as the transmitted data bit \cite{zahedi2000analytical}. Furthermore, we assume that ISI has a negligible effect on the system performance \cite{jamali2015performance}. In this case, conditional error probabilities when bits ``$0$" and ``$1$" are sent can respectively be characterized as follows;
\begin{subequations} \label{P_be}
  \begin{align}
 & \!P_{be}(1|0,l,\vec{\alpha},\bar{H})=\prod_{q=1}^{W}P^{(q)}_{ce-e2e}(1|0,l,\vec{\alpha},\bar{H}),\\
 & \!P_{be}(0|1,l,\vec{\alpha},\bar{H})=1\!-\!\prod_{q=1}^{W}\left[1\!-\!P^{(q)}_{ce-e2e}(0|1,l,\vec{\alpha},\bar{H})\right],\!
  \end{align}
  \end{subequations}
in which $\bar{H}$ is the fading coefficients vector and $\vec{\alpha}=(\alpha_1,\alpha_2,...,\alpha_W)$ is the interference
  pattern occurred on the pulsed mark chips of the first user's OOC, where $\alpha_q$ is the number of interferences on the $q$th pulsed mark chip of the desired user's OOC. Besides, $l$ is the total number of interferences that occurred on all chips of the desired user, i.e., $l=\sum_{q=1}^{W}\alpha_q$.
  Moreover, $P^{(q)}_{ce-e2e}(1|0,l,\vec{\alpha},\bar{H})$ and $P^{(q)}_{ce-e2e}(0|1,l,\vec{\alpha},\bar{H})$ are conditional end-to-end chip error rates on the $q$th transmitted chip of the desired user for ``OFF" and ``ON" states, and can respectively be calculated as follows \cite{jamali2015performance};
  \begin{subequations} \label{p_{ce-e2e}}
     \begin{align}
    & \!\!\!P^{(q)}_{ce-e2e}(1|0,l,\vec{\alpha},{\bar{H}})\!=\!1\!-\!\!\prod_{i=1}^{N+1}\!\left[1\!-\!P^{(q)}_{ce-i}(1|0,l,\vec{\alpha},{\tilde{h}}^{(i)})\right],\!\! \\
    & \!\!\!P^{(q)}_{ce-e2e}(0|1,l,\vec{\alpha},{\bar{H}})\!=\!1\!-\!\!\!\prod_{i=1}^{N+1}\!\left[1\!-\!P^{(q)}_{ce-i}(0|1,l,\vec{\alpha},{\tilde{h}}^{(i)})\right],\!\!
     \end{align}
     \end{subequations}
  where ${\tilde{h}}^{(i)}$ is the $i$th hop fading coefficient and $N$ is the number of intermediate relays. Based on Eqs. \eqref{P_be} and \eqref{p_{ce-e2e}}, we should first determine the chip error rate (CER) of each intermediate hop for ``OFF" and ``ON" states, i.e.,  $P^{(q)}_{ce-i}(1|0,l,\vec{\alpha},{\tilde{h}}^{(i)})$ and $P^{(q)}_{ce-i}(0|1,l,\vec{\alpha},{\tilde{h}}^{(i)})$, respectively.
  
  Since chip detection process during uplink transmission to the first relay is affected by MAI, CER analysis for the uplink transmission to the first relay differs from the other hops CER analysis. Assuming the negligibility of signal-dependent shot noise and considering all users with the same chip power of $P_c$ and all additive noise components, i.e., background light, dark current and thermal noise with Gaussian distribution \cite{jamali2015ber,lee2004part}, the integrated current of the first relay's receiver during uplink transmission can be expressed as follows;
  \begin{align}\label{y1}
  \overrightarrow{y_1}&=\left(y_1^{(1)},...,y_1^{(q)},...,y_1^{(W)}\right)\nonumber\\
  &=RP_cT_c\left[\tilde{h}^{(1)}_{1,1}b^{(1)}_0{L}^{(1)}_{1,1}\overrightarrow{u_1}+{\overrightarrow{\beta}}^{(I)}\right]+\overrightarrow{v_1},
  \end{align}
in which $T_c$ is the chip duration time and $R=\eta q/hf$ is the photodetector's responsivity, where $\eta$, $q$, $h$, and $f$ are the photodetector's quantum efficiency, electron's charge, Planck's constant and the optical frequency, respectively. $\tilde{h}^{(1)}_{1,1}$ and $L_{1,1}^{(1)}$ are the fading coefficient and the aggregated channel loss (due to absorption and scattering effects) of the first hop from the desired user to the first relay, respectively; and $\vec{u}=(1,1,...,1)$ is a $W$-dimensional all-one vector. Moreover, ${\overrightarrow{\beta}}^{(I)}$ is a vector with length $W$ where its $q$th element $\beta_q^{(I)}$ is the weighted sum of $\alpha_q$ independent log-normal RVs, corresponding to the sum of interfering users' fading coefficients, i.e.,
 $\beta_q^{(I)}=\sum_{n\in{\Lambda_q}}L_{n,1}^{(1)}\tilde{h}_{n1}^{(1)}$
 in which $\tilde{h}_{n1}^{(1)}$ and $L_{n,1}^{(1)}$ are respectively the fading coefficient and the aggregated channel loss of the first hop from the $n$th user to the first relay, and ${\Lambda_q}$ specifies the set of $\alpha_q$ interfering users on the $q$th chip. Additionally, $\overrightarrow{v_1}=\left(v_1^{(1)},...,v_1^{(q)},...,v_1^{(W)}\right)$ is a vector with $W$ uncorrelated Gaussian distributed elements each with mean zero and variance $\sigma^2_{T_c}$ corresponding to the integrated combined noise components over $T_c$ seconds \cite{jamali2015ber}. 
 
 Assuming the availability of perfect channel state information (CSI)\footnote{This is a feasible assumption due to the large coherence time of the channel (on the order of $10^{-5}$ to $10^{-2}$ seconds \cite{tang2013temporal}).}, the receiver adopts its threshold value as $\Theta_T=RP_cT_c\tilde{h}^{(1)}_{1,1}{L}^{(1)}_{1,1}/2$ for chip detection process. Therefore, the first hop's CER during uplink transmission can be obtained as follows;
 \begin{align}
& P^{(q,ul)}_{ce-1}(1|0,l,\vec{\alpha},\tilde{h}^{(1)}_{11},\beta_q^{(I)})\nonumber\\
&~~~~~~~~~~~=\Pr\left(y^{(q)}_1>\Theta_T|b_0^{(1)}=0,l,\vec{\alpha},\tilde{h}^{(1)}_{11},\beta_q^{(I)}\right)\nonumber\\
&~~~~~~~~~~~=\Pr\left(v^{(q)}_1>\Theta_T-RP_cT_c\beta_q^{(I)}|l,\vec{\alpha},\tilde{h}^{(1)}_{11},\beta_q^{(I)}\right)\nonumber\\
&~~~~~~~~~~~=Q\left(\frac{RP_cT_c\left[\tilde{h}^{(1)}_{1,1}{L}^{(1)}_{1,1}/2-\beta_q^{(I)}\right]}{\sigma_{T_c}}\right),\label{pce1_0}
\end{align}
 \begin{align}
  & P^{(q,ul)}_{ce-1}(0|1,l,\vec{\alpha},\tilde{h}^{(1)}_{11},\beta_q^{(I)})\nonumber\\
 &~~~~~~~~~~~=Q\left(\frac{RP_cT_c\left[\tilde{h}^{(1)}_{1,1}{L}^{(1)}_{1,1}/2+\beta_q^{(I)}\right]}{\sigma_{T_c}}\right),\label{pce1_1}
  \end{align}
where $Q\left(x\right)=({1}/{\sqrt{2\pi }})\int^{\infty }_x{{\rm exp}({-{y^2}/{2}})}dy$ is the Gaussian-Q function.

For the other hops, either during uplink or downlink transmission, MAI does not affect the chip detection process. Moreover, MAI of the synchronous downlink  transmission can be eliminated when the number of concurrent users satisfy the condition $M<\frac{F}{W^2}+1$ \cite{ghaffari2008wireless}. Therefore, CERs of the remaining hops of uplink transmission as well as all hops of downlink transmission can similar to Eqs. \eqref{pce1_0} and \eqref{pce1_1} be obtained as follows;
 \begin{align}
P^{({\rm MAI-free})}_{ce-q}(1|0,\tilde{h}^{(i)})&=P^{({\rm MAI-free})}_{ce-q}(0|1,\tilde{h}^{(i)})\nonumber\\
&=Q\left(\frac{RP_cT_c\tilde{h}^{(i)}{L}^{(i)}}{2\sigma_{T_c}}\right),\label{pce}
  \end{align}
in which ${L}^{(i)}$ is the $i$th hop aggregated channel loss. Finally, \eqref{pce1_0}-\eqref{pce} can be applied to \eqref{p_{ce-e2e}} to obtain the end-to-end conditional CERs and the result can then be applied to \eqref{P_be} to achieve both the up-and downlink end-to-end conditional BERs of relay-assisted UW-OCDMA network. Moreover, the final BER can similar to \cite{jamali2015performance} be obtained by averaging over fading coefficients and interfering patterns.
 
 \begin{figure}
  \centering
   \includegraphics[width=3.6in]{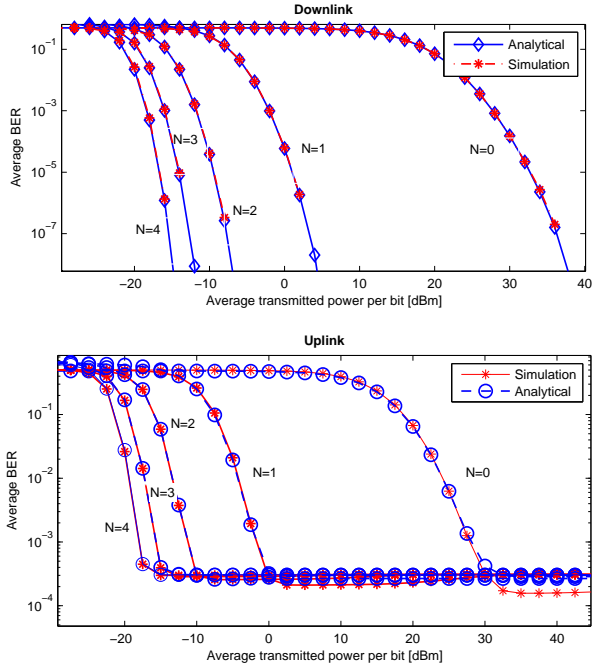}
   \caption{Analytical and simulation results for uplink and downlink BER of the underwater optical network employing multi-hop transmission in clear ocean channel. $N$ denotes the number of equidistant intermediate relays and simulation parameters are chosen as follows: extinction coefficient $c=0.151\ \rm m^{-1}$, number of users $=5$, end-to-end distance $r_0=90 \ \rm m$, bit rate $R_b=2 \ \rm Mbps$, OOC code length $F=50$, OOC code weight $W=3$, OOC's maximum cross-and autocorrelation $\rho = 1$, log-amplitude variance of fading $\sigma_x^2=0.17$ (in a $90\ \rm m$ link) and effective diameter of receiver aperture $D_0=20\ \rm cm$. }
 \end{figure}
Figure 4 shows the the end-to-end BER of relay-assisted UW-OCDMA network for both up-and downlink transmissions. As it can be seen, MAI limits the uplink transmission BER to a predetermined bound \cite{jamali2015performance} while the absence of MAI allows the downlink BER to monotonically decrease with increases on the transmitted power. Moreover, beneficial application of multi-hop transmission is obvious from theses figures; as the number of intermediate relays increases the end-to-end system performance considerably improves.
\subsubsection{Distributed underwater optical sensor networks}
Distributed sensor network is a collection of mobile and fixed sensors each of which has sensing, receiving, transmitting and computing capabilities. Such networks are capable of self-deployment; \textit{i.e.}, starting from some compact initial configuration, the nodes in the network can spread out such that the area covered by the network is maximized. In this scheme, the OOC-encoded data produced by a source sensor is relayed with intermediate sensors until it reaches the OBTS. Deployment of these sensors in our UW-OCDMA network eliminates the need for intermediate relay nodes, but increases the complexity of the sensors' structure.

The cellular UW-OCDMA network can also be designed to be compatible with developed underwater acoustic sensor networks. Underwater acoustic communication suffers from inherently limited bandwidth, severe multi-path and fading, and considerable propagation delay. However, acoustic wave propagates tens of kilometers in underwater medium. Therefore, underwater acoustic sensors can be deployed in our network far away from OBTSs. In this regard, we need to design an acoustic-to-optical converter to make the communication possible.


\subsection{Underwater Localization}
Global positioning system (GPS) receivers are widely used in terrestrial area to determine the location of a mobile user. However, this is not possible in underwater medium since GPS signals do not propagate through water. The infrastructure of the proposed cellular UW-OCDMA network can be used as a promising alternative to GPS; since the OBTSs are placed at pre-determined locations, they can serve as anchor (reference) nodes. According to the range-based localization method, the underwater MU needs to, first, estimate its distances from each OBTS in its communication range. And then estimate its position, using methodologies based on the intersection of various circles centered at each OBTS with radii corresponding to the distance measurements. Several  techniques such as received signal strength (RSS), angle of arrival (AOA), time of arrival (TOA) and time difference of arrival (TDOA) are presented and investigated in terrestrial networks to estimate the MU's distance from the reference nodes \cite{guvenc2009survey}. In the following, we discuss RSS and TDOA as two promising schemes in our UW-OCDMA network and will elaborate RSS technique in more details.

In TDOA technique, the MU's position is estimated based on the difference between TOAs from several OBTSs. Once the MU detects a signal from an OBTS, it waits for another signal from a different OBTS (with different ID) and measures the difference between these two TOAs. As shown in Fig. 3(b), the locations of the points with a constant time difference (constant distance difference) are hyperbolic lines with two OBTSs as the focal points. The differences in receiving time of signals from three nodes result into two independent hyperbolas that their cross point is the location of the MU. Note that this technique requires precise synchronization between all OBTSs and a high resolution clock to measure the time difference.

According to the RSS scheme, the distance is estimated based on the attenuation introduced by the propagation of the signal from OBTSs to MU. As shown in Fig. 3(c), the underwater MU compares the received signals from at least three OBTSs to identify its location. Each level indicates a circle around the corresponding reference node and the cross section of three of such circles specifies the location of the user. Since almost all receivers can estimate the level of received signal, RSS does not need extra devices or modules, which makes it a low cost method. However, a precise channel model is required for an accurate distance estimation using the relation between distance and attenuation behavior. In order to eliminate random nature of the received signal due to the optical channel turbulence, we can average the received signal over a period of time larger than the channel coherence time. 

In the following, the expressions to estimate the first user's position based on RSS scheme are driven. The integrated current at the first user's receiver $y_{(i,1)}$ is considered as a measure of signal strength;
\begin{align}
y_{(i,1)}=RP_{t,avg}T_s{\tilde{h}}_{\left(i,1\right)}L(d_{\left(i,1\right)})+v_1 ,
\end{align}
where $R=\eta q/hf$ is the photodetector's responsivity. $P_{t,avg}$ and $T_s$ denote average power and time duration of the localization signal transmitted by OBTS, respectively. ${\tilde{h}}_{\left(i,1\right)}$ and $L(d_{\left(i,1\right)})$ represent log-normal fading coefficient and the aggregated channel loss (due to absorption and scattering effects) associate with the channel between the $i$th OBTS and the first user, respectively. $v_1$ is a zero mean Gaussian random variable corresponding to the integrated combined noise components over $T_s$ seconds at the first user's receiver.

The above expression shows the integrated current at first user's receiver result from $i$th OBTS localization signal as a function of distance between user and OBTS, $d_{\left(i,1\right)}$. However, in RSS method, we need to obtain user's distance in terms of the received signal strength. Having known some pairs of ($y_{(i,1)}$, $d_{\left(i,1\right)}$) from Monte Carlo simulation or experimentally measurements, a polynomial function can be fitted to obtain an estimate of $d_{\left(i,1\right)}$in terms of a given $y_{(i,1)}$ as follows;
\begin{align} \label{h_hat}
{\hat{d}}_{\left(i,1\right)}=b_0+b_1.y_{(i,1)}+b_2.{y_{(i,1)}}^2+\dots +b_M.{y_{(i,1)}}^M ,
 \end{align}
in which coefficients $\left(b_0, b_1, \dots , b_M\right)$ can be found using minimum mean square error (MMSE) method. 
Therefore, based on the estimated user's distances from neighboring OBTSs $\{{\hat{d}}_{\left(i,1\right)}\}^{N_{neig}}_{i=1}$, we need to approximate the user's position ($x_u,y_u$) in a two dimensional coordinate. For the sake of simplicity, it is assumed that the first OBTS has been located at the origin of the coordinate, i.e., ($0,0$), and the rest of OBTSs are considered to be located at ($x_i,y_i$). Note that we can always move the origin of the coordinator to the first OBTS position to satisfy this assumption. Therefore, the actual distance of the first user from OBTSs will be;
\begin{align}
{d_{\left(1,1\right)}}^2&=x^2_u+y^2_u 
{d_{\left(i,1\right)}}^2\nonumber\\
&={\left(x_i-x_u\right)}^2+{\left(y_i-y_u\right)}^2,~~~~~ i=2,3,\dots ,\ N_{neig}
\end{align}
By subtracting $d_{\left(1,1\right)}$ from $d_{\left(i,1\right)}$, we have;
\begin{align}
&{d_{\left(i,1\right)}}^2-{d_{\left(1,1\right)}}^2\nonumber\\
&~~~=x^2_i+y^2_i-2x_ix_u-2y_iy_u,~~i=2,3,\dots ,\ N_{neig}
\end{align}
Then, defining ${{\rm r}}^2_{i\ }=x^2_i+y^2_i$, the above expression can be written in a matrix form as $\textbf{C}x=\textbf{D}$, where;
\begin{align}
\textbf{C}=\left( \begin{array}{cc}
x_2 & y_2 \\ 
x_3 & y_3 \\ 
\vdots  & \vdots  \\ 
x_{N_{neig}} & y_{N_{neig}} \end{array}
\right),
\end{align}
\begin{align}
\textbf{D}=\frac{1}{2}\left( \begin{array}{c}
{{\rm r}}^2_2\ \ \ -\ \ {d_{\left(2,1\right)}}^2\ \ +\ \ {d_{\left(1,1\right)}}^2 \\ 
 \begin{array}{c}
{{\rm r}}^2_3\ \ -\ \ {d_{\left(3,1\right)}}^2\ \ +\ \ {d_{\left(1,1\right)}}^2 \\ 
\vdots  \\ 
{{\rm r}}^2_{N_{neig}}-{d_{\left(N_{neig},1\right)}}^2+{d_{\left(1,1\right)}}^2 \end{array}
 \end{array}
\right),
\end{align}
\begin{align}
\textit{x}=\left( \begin{array}{c}
x_u \\ 
y_u \end{array}
\right),
\end{align}
\noindent Therefore, substituting $d_{\left(i,1\right)}$ with ${\hat{d}}_{\left(i,1\right)}$, the linear-least-square (LLS) solution for the estimated position of the first user can be written as;
\begin{align} \label{x_hat}
\hat{x}={({{\mathbf C}}^T{\mathbf C})}^{-1}{{\mathbf C}}^T{\mathbf D},
\end{align}
From this matrix form, it is implied that in order to estimate the user's position, we need to know the distance of the user from at least three neighboring OBTSs. However, the contribution of more OBTSs will result in less estimation error.

\begin{figure}
 \centering
  \includegraphics[width=3.6in]{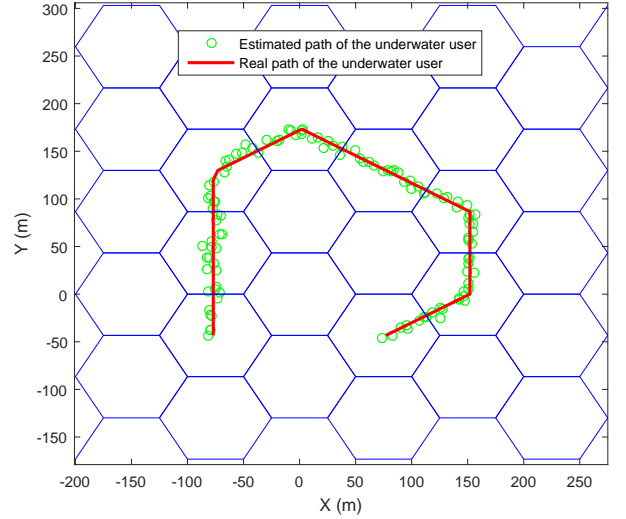}
  \caption{Simulation results of the proposed underwater localization based on RSS algorithm. The Monte Carlo simulation is used to model a pure sea water channel with extinction coefficient of $c=0.043\ \rm m^{-1}$. The cell radius and log-amplitude variance of fading are considered to be $r_0=50$ m and $\sigma_x^2=0.1$, respectively.}
\end{figure}

Figure 5 shows the simulation results for positioning accuracy of the proposed underwater localization based on RSS algorithm in a pure sea water. In this simulation, the underwater mobile user utilizes the expression in \eqref{h_hat} in order to estimate its distance from its seven neighboring OBTSs ($N_{neig}=7$) based on the received signal of these OBTSs. Then, user applies the linear least square estimation in \eqref{x_hat} to approximate its position. Note that in Fig. 5, the deviation from the actual position of the user is result from the random nature of the turbulence-induced fading and additive noise.

\section{Possible Design Challenges in Cellular UW-OCDMA Networks}
In this section, we primarily discuss possible challenges in implementing cellular UW-OCDMA networks such as blockage avoidance, cell-edge coverage, power control, and limitation in total number of users. Then, we present appropriate solutions to address these challenges. 

\begin{figure}
\centering
\includegraphics[width=3.6in]{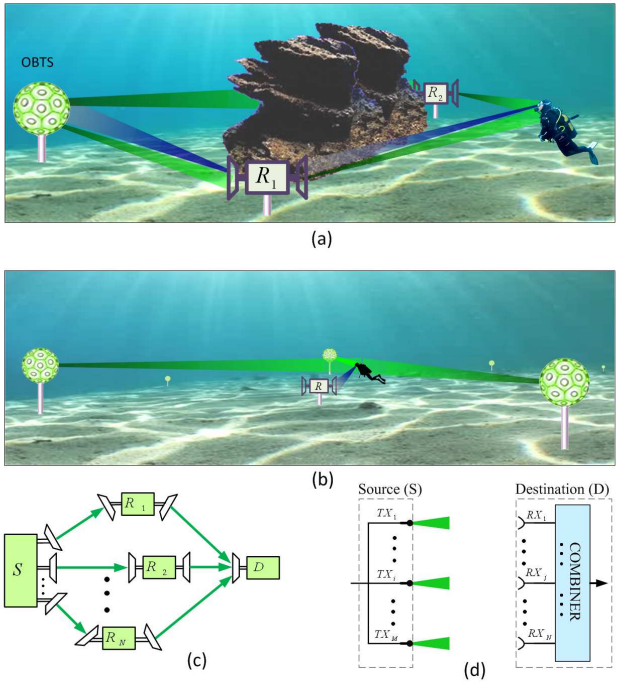}
\caption{(a) Blockage avoidance scenario; (b) cell-edge coverage concept; (c) parallel relaying; (d) spatial diversity.}
 \end{figure}

 \subsection{Blockage Avoidance}
 In certain underwater areas, we may need real-time communication, while there is not any line of sight (LOS) link to any of OBTSs. In such cases, relay nodes can be employed with comparatively simple infrastructure to provide a reliable communication link between MUs and the nearby OBTS, as shown in Fig. 6(a). During uplink transmission, the MU transmits its signal to relay $R_1$, which is located in LOS with both OBTS and MU. The relay $R_1$ applies specific processes (e.g., amplification, detection, or decoding) on the received signal, and then forwards the processed signal toward the OBTS. Reverse strategy can be adopted for downlink transmission. Furthermore, to achieve better performance, more relay nodes (e.g. $R_2$ in Fig. 6(c)) can be employed to form a parallel relaying configuration. In this scheme, all relays can receive the same signal from OBTS and transmit them to the MU; or when channel state informations (CSI) are available at the OBTS, the relay with the most reliable channel can be selected. 
 
 \subsection{Cell-Edge Coverage}
 One of the most important issues in designing a cellular network is how the network can cover cell-edge users suffering from low received signal to noise ratios (SNRs). A promising approach is that all the neighboring OBTSs, which have relatively the same distance to the cell-edge MU, simultaneously transmit optical signal to the MU in order to form a multiple-input single-output (MISO) configuration (see Fig. 6(b)). Since different links can be considered as independent parallel channels, deployment of multiple transmitters (as shown in Fig. 6(d)) can substantially improve the system performance, particularly for highly turbulent channels. However, it requires perfect synchronization strategy among OBTSs to transmit data with appropriate time delays.
 
It can be shown the conditional BER of multiple-input multiple output (MIMO) UWOC system with equal gain combiner can be expressed as \cite{jamali2015ber};
 \begin{align} \label{eq11}
& P^{(\rm MIMO)}_{be|b_0,{\boldsymbol{\bar{H}}},b_k}=\nonumber\\
&
 ~~~~~~Q\left(\frac{{\underset{j=1}{\overset{N_r}{\sum}}}{\underset{i=1}{\overset{N_t}{\sum}}}{\tilde{h}}_{ij}\gamma^{(s)}_{i,j}-(-1)^{b_0}{\underset{j=1}{\overset{N_r}{\sum}}}{\underset{i=1}{\overset{N_t}{\sum}}}{\tilde{h}}_{ij}{\underset{k=-L_{i,j}}{\overset{-1}{\sum}}}2b_k\gamma^{(k)}_{i,j}}{2\sqrt{N_r}\sigma_{T_b}}\right),
 \end{align}
 in which $N_t$ and $N_r$ are the number of transmitters and receivers, respectively. ${\boldsymbol{\bar{H}}}=({\tilde{h}}_{11},...,{\tilde{h}}_{N_tN_r})$ is the fading coefficients vector in an $N_t\times N_r$ MIMO system and $\{b_k\}_{k=-L_{ij}}^{0}$ is the transmitted data sequence, where $L_{ij}$ interprets the memory of the channel between the $i$th transmitter and the $j$th receiver. Each fading coefficient is assumed with log-normal distribution, i.e., ${\tilde{h}}_{ij}=\exp(2X_{ij})$ where $X_{ij}$ has a Gaussian distribution with mean $\mu_{X_{ij}}$ and variance $\sigma^2_{X_{ij}}=-\mu_{X_{ij}}$ \cite{gerccekciouglu2014bit,andrews2005laser}.
 Moreover, $\gamma^{(s)}_{i,j}={R}\int_{0}^{T_b}\Gamma_{i,j}(t)dt$ and $\gamma^{(k)}_{i,j}={R}\int_{0}^{T_b}\Gamma_{i,j}(t-kT_b)dt={R}\int_{-kT_b}^{-(k-1)T_b}\Gamma_{i,j}(t)dt$, where $T_b$ is the bit duration time, $\Gamma_{i,j}(t)=P_i(t)*h_{0,ij}(t)$ is the received optical signal from the $i$th transmitter to the $j$th receiver, $P_i(t)$ is the transmitted optical pulse shape of the $i$th transmitter, $h_{0,ij}(t)$ is the fading-free impulse response of the channel between the $i$th transmitter and the $j$th receiver (obtained from Monte Carlo simulations, in a similar approach to \cite{tang2014impulse}), and $*$ represents the convolution operator. Additionally, $\sigma^2_{T_b}$ is the variance of the zero mean Gaussian distributed integrated current of the $j$th receiver over each $T_b$ seconds, corresponding to the combined noise components of the receiver \cite{jamali2015ber}.
Assuming the maximum channel memory as $L_{\rm max}=\max\{L_{11},L_{12},...,L_{N_tN_r}\}$, the final BER of MIMO-UWOC system can be obtained by averaging over ${\boldsymbol{\bar{H}}}$ (through an ($N_t\times N_r$)-dimensional integral) as well as averaging over all $2^{L_{\rm max}}$ sequences for $b_k$s;
\begin{align}\label{eq13}
& P^{(\rm MIMO)}_{be}\nonumber\\
&~~~=\frac{1}{2^{L_{\rm max}}}\sum_{b_k}\int_{{\boldsymbol{\bar{H}}}}\frac{1}{2}\left[P^{(\rm MIMO)}_{be|1,{\boldsymbol{\bar{H}}},b_k}+P^{(\rm MIMO)}_{be|0,{\boldsymbol{\bar{H}}},b_k}\right]f_{{\boldsymbol{\bar{H}}}}({\boldsymbol{\bar{H}}})d{\boldsymbol{\bar{H}}},
\end{align}
where $f_{{\boldsymbol{\bar{H}}}}({\boldsymbol{\bar{H}}})$ is the joint PDF of fading coefficients in ${\boldsymbol{\bar{H}}}$. Furthermore, the ($N_t\times N_r$)-dimensional integral in \eqref{eq13} can effectively be calculated through an ($N_t\times N_r$)-dimensional finite series using Gauss-Hermite quadrature formula \cite{jamali2015ber}.
It is worth mentioning that in the case of negligible ISI the average BER of MIMO-UWOC system simplifies to;
\begin{align}
P^{(\rm MIMO)}_{be,{\rm ISI-free}}=\int_{{\boldsymbol{\bar{H}}}}Q\left(\frac{\sum_{j=1}^{N_r}\sum_{i=1}^{N_t}{\tilde{h}}_{ij}\gamma^{(s)}_{i,j}}{2\sqrt{N_r}\sigma_{T_b}}\right)f_{{\boldsymbol{\bar{H}}}}({\boldsymbol{\bar{H}}})d{\boldsymbol{\bar{H}}}.
\end{align}

Performance of the system employing MISO configurations is simulated  for two different values of log-amplitude variance of fading $\sigma_x^2=0.01$ and $\sigma_x^2=0.16$. The simulation results depicted in Fig. 7 show a significant performance improvement by increasing the number of transmitters particularly for more turbulent channels.
\begin{figure}
\centering
\includegraphics[width=3.6in]{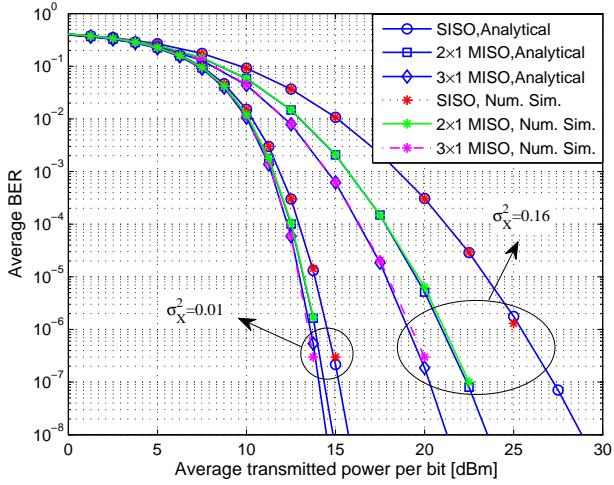}
\caption{Analytical results along with numerical simulations for performance of the system employing MISO configurations in coastal water channel; extinction coefficient $c=0.398 \ \rm m^{-1}$, communication range $r_0=25\ \rm m$, and transmission rate $R_b=1~{\rm Gbps}$. }
 \end{figure}
\subsection{Power Control}
Apart from the previous challenges, power consumption needs to be insightfully considered in designing the proposed UW-OCDMA network. MUs and OBTSs often use limited power supplies or need to reduce their corresponding costs. Power allocation is an important issue in underwater medium because this environment absorbs more energy in comparison with free space or fiber optics medium. Furthermore, increasing power beyond safety standards may harm underwater ecosystem. 

In the first presentation of our proposed cellular UW-OCDMA in \cite{akhoundi2014cellular}, we assumed that each OBTS transmits a constant power omnidirectionally, \textit{i.e.}, all LEDs mounted on OBTS are turned on to send information to  MUs, regardless of the MUs' position in the cell. Moreover, the power allocated to each user was the same for all MUs, no matter how the channel quality of each user is (see Fig. 8(a)). However, this scheme is not  efficient in terms of total energy consumption. To reduce the network required power, two algorithms are proposed in this section.
\begin{figure}
 \centering
  \includegraphics[width=3.4in]{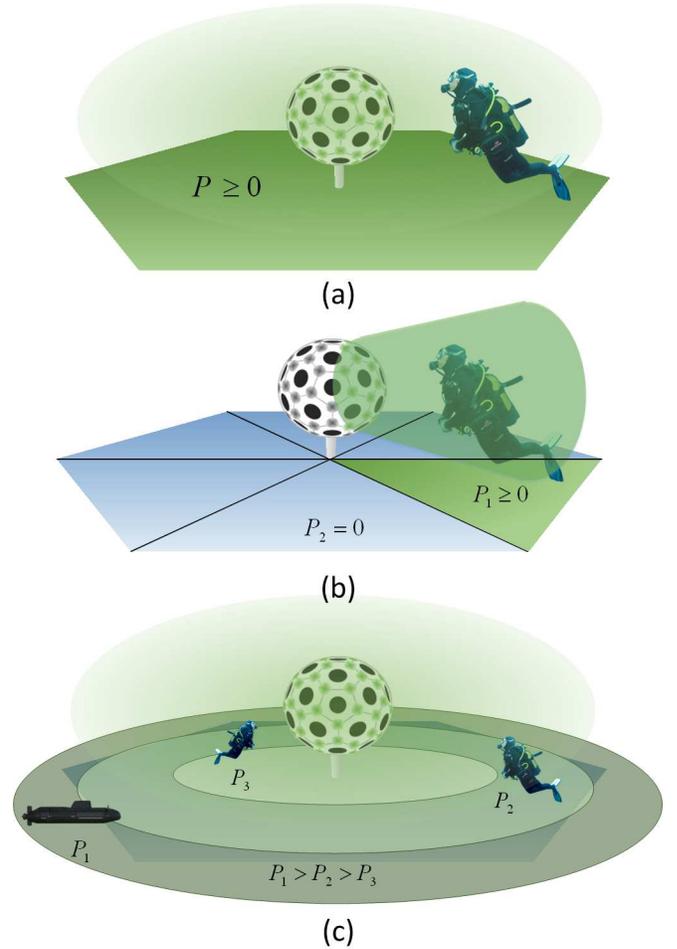}
  \caption{Different power control schemes: (a) omnidirectional transmission; (b) partitioning the cell into sectors, where the number of sectors is $N_S=6$; (c) allocating power according to the users' channel quality, where the number of rings is $N_R=3$.}
\end{figure}
 \subsubsection{Partitioning the cell into sectors}
In this algorithm, as shown in Fig. 8(b), we divide the LEDs and photodiodes mounted on OBTS into $N_{s}$ sectors and only one of these sectors will be active to communicate with each MU. To realize this scheme, the OBTS needs to know corresponding sectors in the cell in which each MU is located. For this purpose, each MU sends a beacon message towards OBTS in order to request for communication. The OBTS determines the MU's sector considering the fact that beacon message was received from which OBTS sections. As the user moves in the cell, location of the MU will be updated and consequently active section of OBTS will change.
 \subsubsection{Allocating power according to the user's channel quality}
In this algorithm, instead of assigning equal power to all MUs within the cell regardless of their channel quality, we can utilize MU's quantized channel information (QCI) to control the transmitting power and keep the bit error rate on a desired value in several points in the cell, as it is shown in Fig. 8(c). According to the MUs' distances to the OBTS and their channels quality, they will be divided into $N_R$ rings and we allocate appropriate power to different rings. The advantage of using QCI instead of full CSI is that we can avoid enormous complexity of instantaneous calculations at OBTSs. The ability to change the transmission power of each user will also allow us to allocate power according to the requested service. For instance, those users requesting voice services, consume less power than those requesting video services. 

\begin{figure}
 \centering
  \includegraphics[width=3.6in]{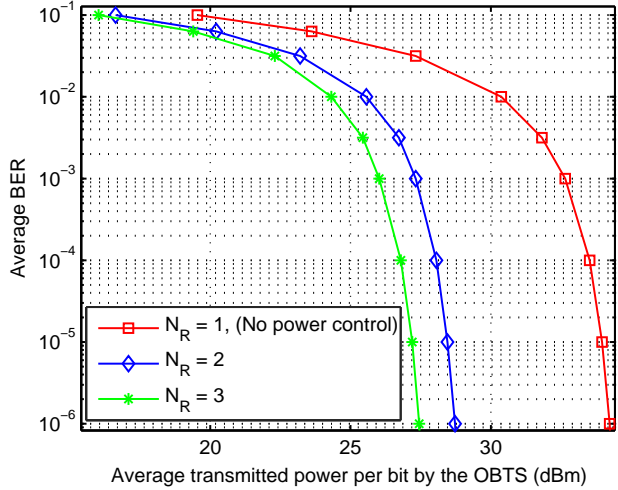}
  \caption{Simulated downlink performance of the system employing the second power control algorithm in clear ocean channel. It is assumed that MUs are distributed uniformly in the cell. Simulation parameters are chosen as follows: extinction coefficient $c=0.151\ \rm m^{-1}$, number of users $=5$, cell radius $r_0=90 \ \rm m$, OOC code length $F=50$, OOC code weight $W=3$, OOC's maximum cross-and autocorrelation $\rho = 1$, log-amplitude variance of fading depends on MUs' distance from the OBTS and varies between $\sigma_x^2=0\sim0.14$ and effective diameter of receiver aperture $D_0=20\ \rm cm$. }
\end{figure}
The performance analysis of these two algorithms with details is explored in \cite{banihassan2015adaptive}. However, in Fig. 9, the second algorithm performance is demonstrated in terms of average transmitted power per bit. Three different schemes in Fig. 9 are as follows: allocating equal power to all MUs ($N_R=1$), dividing MUs to two, and three rings with respect to their channel quality ($N_R=2$ and $N_R=3$). The simulation results show a near $6 dB$ gain in ${\rm BER}=10^{-6}$ when employing power control algorithm with $N_R=3$.

\subsection{Expanding Number of Supported Users}
In the proposed cellular UW-OCDMA network, the capacity of each OBTS, in terms of the number of covered MUs, depends on the number of OOC codes ($N_c$) in OCDMA system. There are two approaches for code assignment at the wireless front end, unique assignment of OOCs over the network or reusing OOCs in OBTSs. In the first approach, there is not any interference between neighboring OBTSs, while in the second one, neighboring OBTSs might interfere together, if they use the same code set. Although in the first approach OBTSs do not interfere together, the capacity of the network is limited to $N_c$. Thus, reusing OOCs is an attractive solution to increase the capacity of underwater cellular network, provided that we resolve its interference issue.

In order to mitigate interference among neighboring OBTSs, OOCs are divided into three subsets, $S_1, S_2,$ and $S_3$, and these subsets are assigned to OBTSs such that neighboring OBTSs have different OOC sets. Fig. 10(a) illustrates the proposed OOCs assignment. In this approach, the capacity of cellular network is $\frac{N_c}{3}×N_{\rm OBTS}$, where $N_{\rm OBTS}$ denotes the number of OBTSs.
 \begin{figure}
 \centering
  \includegraphics[width=3.6in]{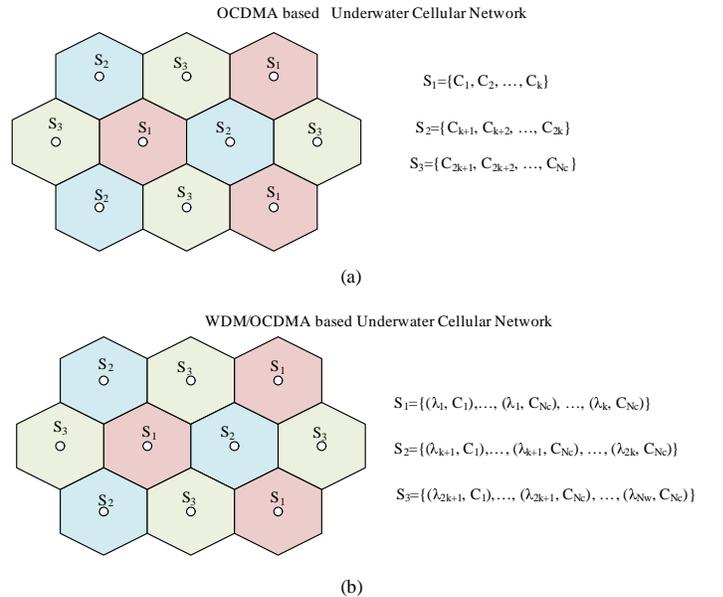}
  \caption{(a) Code reuse in OCDMA and; (b) Wavelength/code reuse in WDM/OCDMA-based underwater cellular network`.}
\end{figure}
The number of channels at the wireless front end can be further increased by integrating OCDMA with wavelength division multiplexing (WDM) scheme (WDM/OCDMA). Let $N_w$ denote the number of available wavelengths in WDM system, then $N_w$ wavelengths are divided into three subsets, whereby the subsets are assigned to neighboring OBTSs. In Fig. 10(b), the channel assignment in the proposed WDM/OCDMA-based underwater cellular network is depicted. We note that by using WDM technique the capacity of underwater cellular network is increased to $\frac{N_w}{3}×N_c×N_{\rm OBTS}$. It is worth mentioning that in OCDMA-based system wireless channels are identified with OOCs ($C_i$), whereas in WDM/OCDMA counterpart each channel is specified with wavelength/OOC pair ($\lambda_j, C_i$). 

Although WDM/OCDMA-based system provides higher capacity, it needs more complicated hardware in both OBTS and MU. Instead of single band transceivers employed in the OCDMA scheme, in WDM/OCDMA counterpart, multi-wavelength transceivers are required. 
\section{Conclusions}
In this paper, an overview of characteristics, challenges and potentials of the cellular underwater wireless optical code division multiple-access (UW-OCDMA) network based on optical orthogonal codes (OOC) are described. The primary aim of this network is to define a flexible, reliable and practical framework to satisfy the military and commercial  demands of the underwater communication.  In the cellular UW-OCDMA, a set of optical base transceiver stations (OBTSs) are placed at the center of hexagonal cells to cover a large underwater area. All OBTSs can be linked via fiber optic to a central node namely optical network controller (ONC). Furthermore, additional applications of the proposed network in relay-assisted underwater local sensor networks and underwater localization and positioning are presented.  Finally, probable design challenges regarding blockage avoidance, cell-edge coverage, power control and limitation in network capacity are discussed and proper strategies which can be adopted to overcome these issues are addressed.
\end{document}